\documentclass[]{emulateapj}
\usepackage{CJKutf8}
\usepackage{apjfonts}

\shorttitle{Surface Brightness and Mass-metallicity Relation}

\begin{document}

\title{The influence of galaxy surface brightness on the mass-metallicity relation }

\author{Po-Feng Wu \begin{CJK*}{UTF8}{bsmi}(吳柏鋒)\end{CJK*} \altaffilmark{1}, Rolf-Peter Kudritzki\altaffilmark{1,2}, R. Brent Tully\altaffilmark{1}, and J. D. Neill\altaffilmark{3} }

\altaffiltext{1}{University of Hawaii, Institute for Astronomy, 2680 Woodlawn Drive, HI 96822, USA}
\altaffiltext{2}{University Observatory Munich, Scheinerstr. 1, D-81679 Munich, Germany}
\altaffiltext{3}{California Institute of Technology, 1200 E. California Blvd. MC 278-17, Pasadena, CA 91125, USA}

\begin{abstract}

We study the effect of surface brightness on the mass-metallicity relation using nearby galaxies whose gas content and metallicity profiles are available. Previous studies using fiber spectra indicated that lower surface brightness galaxies have systematically lower metallicity for their stellar mass, but the results were uncertain because of aperture effect. With stellar masses and surface brightnesses measured at WISE W1 and W2 bands, we re-investigate the surface brightness dependence with spatially-resolved metallicity profiles and find the similar result. We further demonstrate that the systematical difference cannot be explained by the gas content of galaxies. For two galaxies with similar stellar and gas masses, the one with lower surface brightness tends to have lower metallicity. 
Using chemical evolution models, we investigate the inflow and outflow properties of galaxies of different masses and surface brightnesses. We find that, on average, high mass galaxies have lower inflow and outflow rates relative to star formation rate. On the other hand, lower surface brightness galaxies experience stronger inflow than higher surface brightness galaxies of similar mass. The surface brightness effect is more significant for low mass galaxies. 
We discuss implications on the different inflow properties between low and high surface brightness galaxies, including star formation efficiency, environment and mass assembly history.  

\end{abstract}


\section{Introduction}

Chemical enrichment is one of the keys to understand the evolution of galaxies. Heavy elements, or metals, are synthesized in stars then released into the interstellar medium (ISM) by stellar winds or supernova explosions. The metal-enriched ISM subsequently acts as the raw material for the next generation of new-born stars. Tight relations between the stellar mass and both gas-phase and stellar metallicities were discovered, where galaxies with larger stellar mass on average have higher metallicities \citep{tre04,gal05,sav05,lee06}. 

To first order, the origin of the mass-metallicity relation can be understood as a simple process of recycling of metals in the ISM. Galaxies with larger stellar mass have synthesized and released more heavy element throughout their lives, therefore higher metallicity is naturally expected. Nevertheless, this simple picture does not tell the whole story. As pointed out by \citet{tre04}, the dispersion in the mass-metallicity relation cannot be explained simply by measurement uncertainties of metallicity. There must be other factors affecting the metallicity. 

Several studies have been searching for parameters beyond stellar mass which affects metallicity. An anti-correlation between gas-phase metallicity and star formation rate (SFR) has been found both locally and at high redshifts, where at fixed stellar mass, galaxies with higher SFR have lower metallicity \citep{ell08,man10,lar10,yab15}. 
A similar anti-correlation seems also to be present between the gas content and metallicity, where gas-rich galaxies have on average lower metallicity \citep{bot13,hug13}. There is also a dependency on galaxy structure, where larger, or lower surface brightness, galaxies tend to have lower metallicity for their stellar mass \citep{ell08,lia10,sal14}. 

Most of these results are obtained with data from SDSS fiber spectra. Although the sample size is large and provides good statistics, data from fiber spectra have fundamental issues. The spectroscopic information, including SFR and metallicity, comes from only the center of the galaxy, while stellar mass, gas mass and galaxy structure are derived from imaging with different apertures. Moreover, the coverage varies among galaxies, depending on the size, structure and distance of galaxies. Therefore, some of the dependencies on a second parameter could come from aperture effects \citep{ell08,hug13,san13}. This calls for an approach, where spatially resolved information of metallicity is used.

In the interpretation of the observed relationship between integrated metallicity and stellar mass, it has become clear that a simple closed box chemical evolution model with no exchange of material between the galaxies and their outside environment \citep{sea72,pag75} is insufficient. In reality, galaxies experience gas inflows from mergers and accretion and outflows caused by feedback from supernova, starburst or AGN and, thus the close-box assumption does not hold. Exchanging material between galaxies and the environment modifies the metallicity of galaxies. By introducing inflow and outflow activities, theoretical models are able to reproduce quantitatively the observed integrated mass-metallicity relation \citep{spi10,pee11,zah12,lil13,zah14}.

With spatially resolved observations, to overcome the uncertainties mentioned above a modified theoretical approach is also required, which utilizes the additional information coming from measurement of radially resolved metallicity profiles. Recently, a variety of such models has been presented \citep{pil12,mot13,kud15}. This allows to impose constraints on the inflow and outflow properties of individual galaxies by chemical evolution models \citep{asc15,kud15}, and then to investigate whether there is any dependence on other physical properties of galaxies.

In this paper, we focus on the surface brightness effect on the mass-metallicity relation. The surface brightness, or surface density, of a galaxy depends on the angular momentum of galaxy for a given mass \citep{dal97,mo98}, which is affected by the mass loss and accretion history of a galaxy \citep{dut12}. We will examine the interplay among surface brightness, metallicity, and properties of inflow and outflow of galaxies from nearby galaxies, where the spatially-resolved information is available. 
In Section~2, we describe the data sets used in the analysis. Section~3 presents the dependence of metallicity on surface brightness inferred from both spectroscopy and broadband color in two independent samples. We use chemical evolution models to constrain the inflow and outflow of galaxies in Section~4, and discuss possible implications and systematics in Section~5. The summary is given in Section~6. We use AB magnitudes in this paper unless noted.

\section{Data and Analysis}

\subsection{Samples}

\begin{figure*}[t]
	\includegraphics[width=\textwidth]{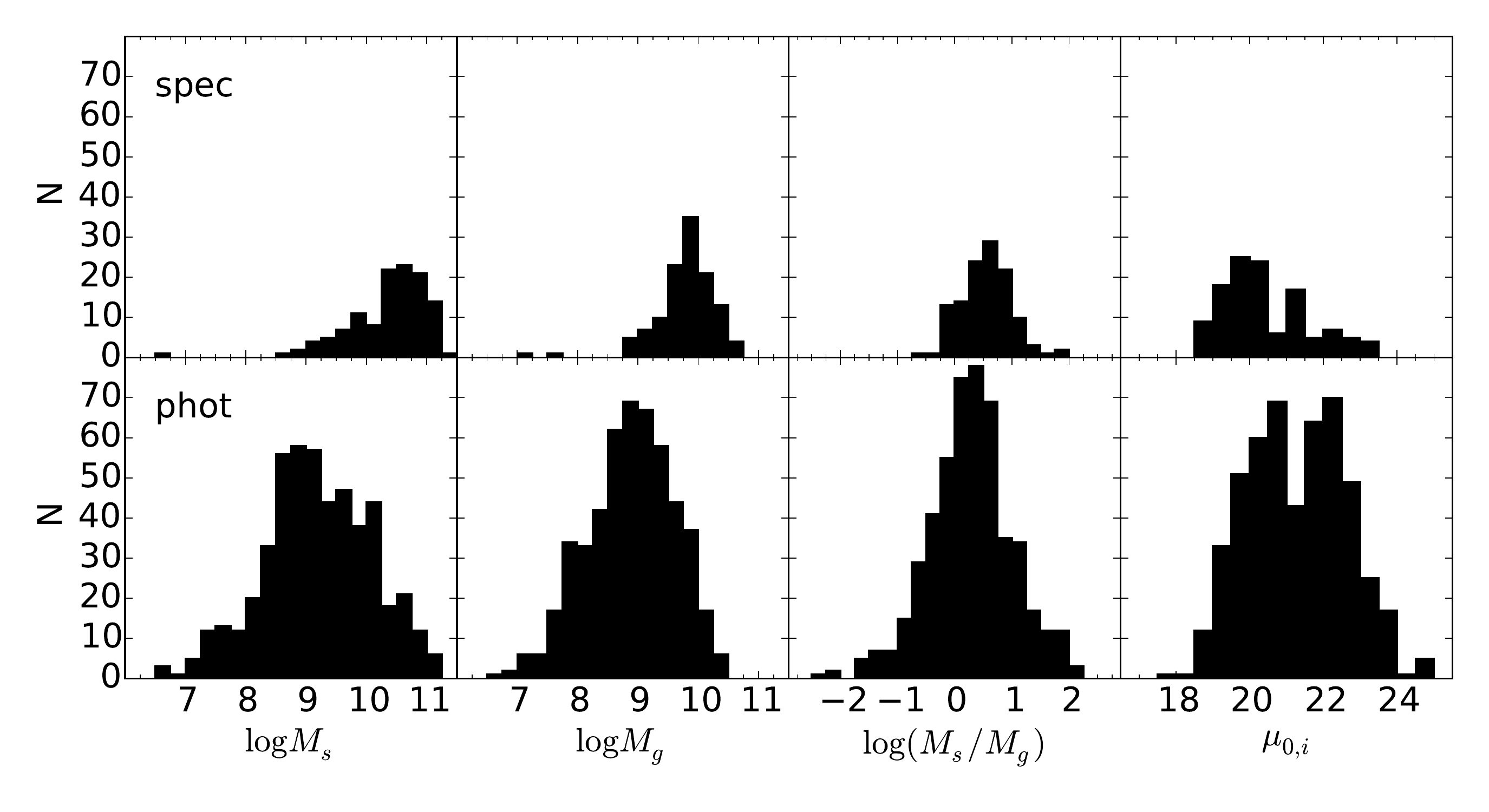}
	\caption{Distributions of stellar mass, gas mass, stellar-to-gas mass ratio and surface brightness of the spectroscopic sample (top) and the photometric sample (bottom).}
	\label{fig:dist}
\end{figure*}

In this paper, we present two samples of galaxies with metallicities estimated from spectra of \ion{H}{2} regions and broadband colors respectively. We use data from surveys including the Wide-Field Infrared Explore \citep[WISE;][]{wri10}, the Sloan Digital Sky Survey \citep[SDSS;][]{yor00}, and the Arecibo Legacy Fast ALFA Survey \citep[ALFALFA;][]{hay11}. The synergy among these large surveys allows us study large number of galaxies with homogeneous observations. The distribution of the stellar mass, gas mass, gas fraction, and surface brightness of the sample is presented in Figure~\ref{fig:dist}.

\subsubsection{Spectroscopic Sample}

The first sample consists of 118 galaxies with gas-phase metallicities measured from \ion{H}{2} regions in \citet{pil14a}. This paper compiled published spectra of \ion{H}{2} regions in 130 nearby galaxies and determined the gas-phase metallicity of each \ion{H}{2} region in a uniform way. For each galaxy, \citet{pil14a} reported the central metallicity and metallicity gradient as a function of radius. 
This set of galaxies provides a homogenized sample with spatially resolved information. We select galaxies whose \ion{H}{1} flux is available for further discussion and refer this sample as the spectroscopic sample

The first row of Figure~\ref{fig:dist} shows the properties of the spectroscopic sample. The majority of galaxies in this sample have stellar mass $M_\ast \gtrsim 10^{10}M_\odot$ and relatively high surface brightness, while the distributions have tails towards lower mass and lower surface brightness end. This sample covers a specific range of stellar-to-gas mass ratio, $0 \lesssim \log(M_\ast/M_g) \lesssim 1$. Few galaxies are located beyond this range.

\subsubsection{Photometric Sample}

The second sample is drawn from the ALFALFA $\alpha$.40 data release \citep{hay11}. Our sample selection starts from ALFALFA sources with heliocentric velocity $V_{hel} \le 3000$ km s$^{-1}$. The velocity, or equivalent, distance limit is chosen as a compromise between sample size and completeness. At larger distances, only HI-massive galaxies will be detected by ALFALFA. Also, the angular resolution of the WISE images prevents us from measuring the structure parameters of stellar disks for fainter or less extended galaxies, thus causing potential bias. 

We then measure the magnitude and surface brightness profile of each ALFALFA source within the velocity limit in the WISE W1 $3.4 \mu$m band. The procedure will be described in Section~\ref{sec:measure}. We select galaxies with $m_{W1} \le 16$, $b \ge 0.8$\arcmin, where b is the minor axis that measured at $\mu_{W1}=25$ mag arcsec$^{2}$, and axis ratio $b/a \ge 0.35$. We consider measurements from galaxies fainter or smaller than these limits are not reliable. The limit on axis ratio removes edge-on galaxies, where correction of the geometric effect of the inclination is less certain. 

Lastly, we excluded galaxies not covered by SDSS imaging. We end up with a sample of 488 galaxies. This sample covers wider ranges of stellar mass, gas mass, stellar-to-gas mass ratio, and surface brightness than the spectroscopic sample, especially for the low mass, low surface brightness and gas-rich end of the distribution (Figure~\ref{fig:dist}). We refer to this sample as the photometric sample. 

\subsection{Photometry and Surface Brightness Profile}
\label{sec:measure}
For this study we perform surface photometry on the WISE W1 and W2 band, and SDSS $u'$, $g'$, and $z'$ band images of each galaxy, using elliptical apertures with fixed centers, orientations and shapes with varying major axes in step of 3\arcsec. Foreground stars and contaminating neighbor galaxies are masked and excluded from measurements. 

We determine ellipses for surface photometry from W1 images as following. We use the shape from the RC3, or SDSS $r'$-band when needed, to measure the sky level for the first pass. We then fit an ellipse to the isophot at 25~mag~arcsec$^{-2}$ of the sky-subtracted image. If the shape and the orientation of the output ellipses are consistent with those of input ellipses, we use it as our input for surface photometry. If the shapes of input and output ellipses are inconsistent ($\Delta \mbox{P.A.} > 20\deg$ or $\Delta b/a > 0.2$), we choose the better one based on visual inspection. 
The chosen orientation and shape are then applied for surface photometry at all bands. We correct the Galactic extinction base on the extinction map of \citet{sch11} and estimate the internal extinction following the prescription of \citet{ver01}. Throughout this paper, we convert extinction among different filters using the conversion factors provided by \citet{yua13}.

We use the central surface brightness of the disk in the W1 band as the surface brightness discussed in this study. The 3.4$\mu$m W1 band is nearly free from concern for internal extinction. We fit the annuli-averaged surface brightness profile in W1 using an exponential form of 
\begin{equation}
\mu(r) = \mu_0 \times \exp (-r/h),
\end{equation}
, where $\mu_0$ is the central surface brightness of the disk and $h$ is the scale length. To exclude the effect of the bulge, we adopt a fiducial fitting range between effective radius, $r_e$, and the radius of isophot of 25.5 mag arcsec$^{-2}$, $r_{25.5}$. In some cases, the surface brightness at $r_e$ is still affected by the bulge or bar. We then manually adjust the inner fitting range to avoid the effect from structures in the inner part of the galaxy. As demonstrated by \citet{mcd09}, we expect little to no statistical difference between the central surface brightness derived from this method and the bulge-disk decomposition.

The central surface brightness is then corrected for the geometric effect of the inclination as follow:
\begin{equation}
\mu_{0,i} = \mu_0 - 2.5 \log(b/a),
\end{equation}
where $a$ and $b$ are the major and minor axis of the galaxy, respectively. The inclination-corrected central surface brightness, $\mu_{0,i}$, is the quantity we use for analysis in this paper.

\subsection{Stellar and Gas Masses}

For stellar masses, we adopt the color-dependent mass-to-light ratio from $W1$ and $W2$ magnitudes \citep{clu14}: 
\begin{equation}
\log \Upsilon = -1.93 (W1_{Vega} - W2_{Vega} ) - 0.04,
\end{equation}
where $W1_{Vega} = W1_{AB} -2.699$, $W2_{Vega} = W2_{AB} - 3.339$, and $\Upsilon$ is the mass-to-light ratio at $W1$ band. $\Upsilon$ is calculated for every annulus until the largest annulus with reliable W2 photometry and then fixed for larger radius. Our sample galaxies have an typical mass-to-light ratio $\Upsilon = 0.57$.

The \ion{H}{1} mass is calculated as $M_{HI} = 2.356 \times 10^5 \times D^2 \times F$, where $D$ is distance in Mpc and $F$ is \ion{H}{1} flux in Jy km s$^{-1}$. We adopt $M_{atom} = 1.4 \times M_{HI}$ to include the contribution from helium and metals. The \ion{H}{1} flux for the spectroscopic sample comes from the Extragalactic Distance Database\footnote{http://edd.ifa.hawaii.edu/} \citep[EDD;][]{tul09}. For the photometric sample, we adopt the HI flux from ALFALFA.

We do not have direct observations of the molecular gas towards our galaxies. To estimate the molecular gas content of our galaxies, we refer to the result of \citet{bot14}, who measured CO gas content towards $\sim 100$ galaxies. \citet{bot14} found that, the molecular-to-atomic gas mass ratio ($M_{H2}/M_{HI}$) has a positive correlation with stellar mass of galaxies with some scatter. From figure~2 of \citet{bot14}, we estimate the molecular-to-atomic gas mass ratio:

\begin{equation}
\label{eq:mol}
\log (M_{H2} / M_{HI}) = 0.66 \times \log(M_\ast/M_{\odot}) - 7.392
\end{equation}

Therefore, the molecular mass is estimated from the combination of stellar and \ion{H}{1} mass. Throughout this paper, ''gas mass'' refers to the total gas mass, $M_{atom} + M_{H2}$. We will discuss the uncertainty and possible systematics in Section~\ref{sec:sys}. 

For distances, we use EDD as the main source. We adopt the group distance, Dgp, in the ``Cosmicflows-2 Distances`` section of EDD as our primary distance. For distances not available in EDD, we take distances from the ALFALFA survey and converted to $H_0 = 75 \mbox{ km s}^{-1} \mbox{Mpc}^{-1}$ in order to be consistent with EDD distances.

\section{The Surface Brightness Dependence on Metallicity}	
\label{sec:result}
In this section, we present the correlation between metallicity and surface brightness from two methods in two independent samples. Going one step farther than existing studies, our analysis considers both the stellar mass and gas mass, and shows that the surface brightness dependence does not come from correlations between surface brightness and stellar mass or gas mass alone. 

\subsection{Gas-phase Metallicity from Spectroscopy}

All galaxies in our spectroscopic sample have central metallicity and metallicity gradient reported in \citet{pil14a}. The first step is to verify whether the result in previous studies, that the higher surface brightness galaxies have higher metallicities at given stellar mass, still holds if we take the average metallicity of the whole galaxy instead of the central metallicity as measured from fiber spectra. However, without knowing the spatial distribution of the gas, we cannot directly convert this information into average metallicity over the whole galaxy. Here we estimate the average metallicity by assuming an exponential gas disk, whose scale length scales with the size of the optical disk ($R_{25}$) \citep{big12}:
\begin{equation}
\Sigma_{gas} \propto e^{(-1.65\times r / R_{25})},
\label{eq:model}
\end{equation}
We calculate the average metallicity out to $R_{25}$: 
\begin{equation}
\langle Z \rangle = \frac{\int_{0}^{R_{25}}\Sigma_{gas} \, (Z_0 +  dZ/dr \times r)\, 2\pi r\, dr}{\int_{0}^{R_{25}}\Sigma_{gas}\, 2\pi r\,dr},
\end{equation}
where $\langle Z \rangle$, $Z_0$, and $dZ/dr$ are the average metallicity, the central metallicity and the metallicity gradient, respectively.

\begin{figure*}[th]
	\centering
	\includegraphics[width=0.85\textwidth]{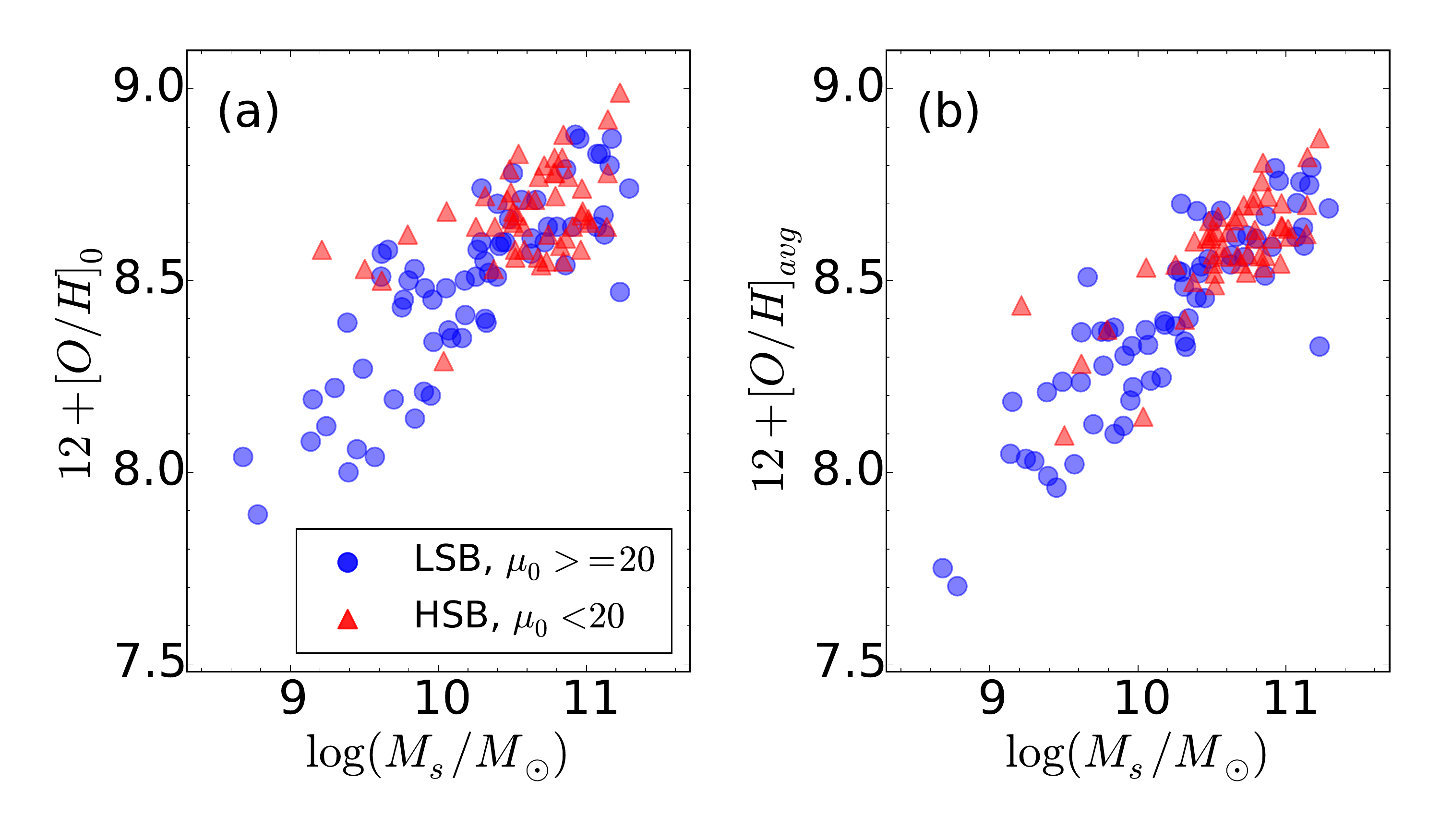}
	\caption{\textit{Left:} Central metallicity as a function of stellar mass, color-coded by surface brightness. The higher and lower surface brightness galaxies are red triangles and blue circles, respectively. The surface brightness dependence is more prominent at low mass end. \textit{Right:} Average metallicity as a function of stellar mass. Higher surface brightness galaxies are mostly located above the mean at low mass, but the difference is not as prominent. }
	\label{fig:mz_spec}
\end{figure*}

Figure~\ref{fig:mz_spec}a shows the central metallicity as a function of the stellar mass of the spectroscopic sample. The sample is divided into two subsample by surface brightness. First of all, the central metallicity shows the mass-metallicity relation as expected. Overall, higher surface brightness galaxies are more metal-rich, mainly due to their more massive nature. But it is also clear that, at $M_\ast \lesssim 10^{10.5} M_{\odot}$, most higher surface brightness galaxies have the highest metallicity for given mass, while no obvious distinction exists between two subsamples at higher mass. This is qualitatively in agreement with previous studies by SDSS fiber spectra \citep{ell08,lia10,sal14}, where for a given mass, high surface brightness, or compact, galaxies have higher metallicities, and the distinction is larger at the low mass end. 

\begin{figure}[th]
	\includegraphics[width=\columnwidth]{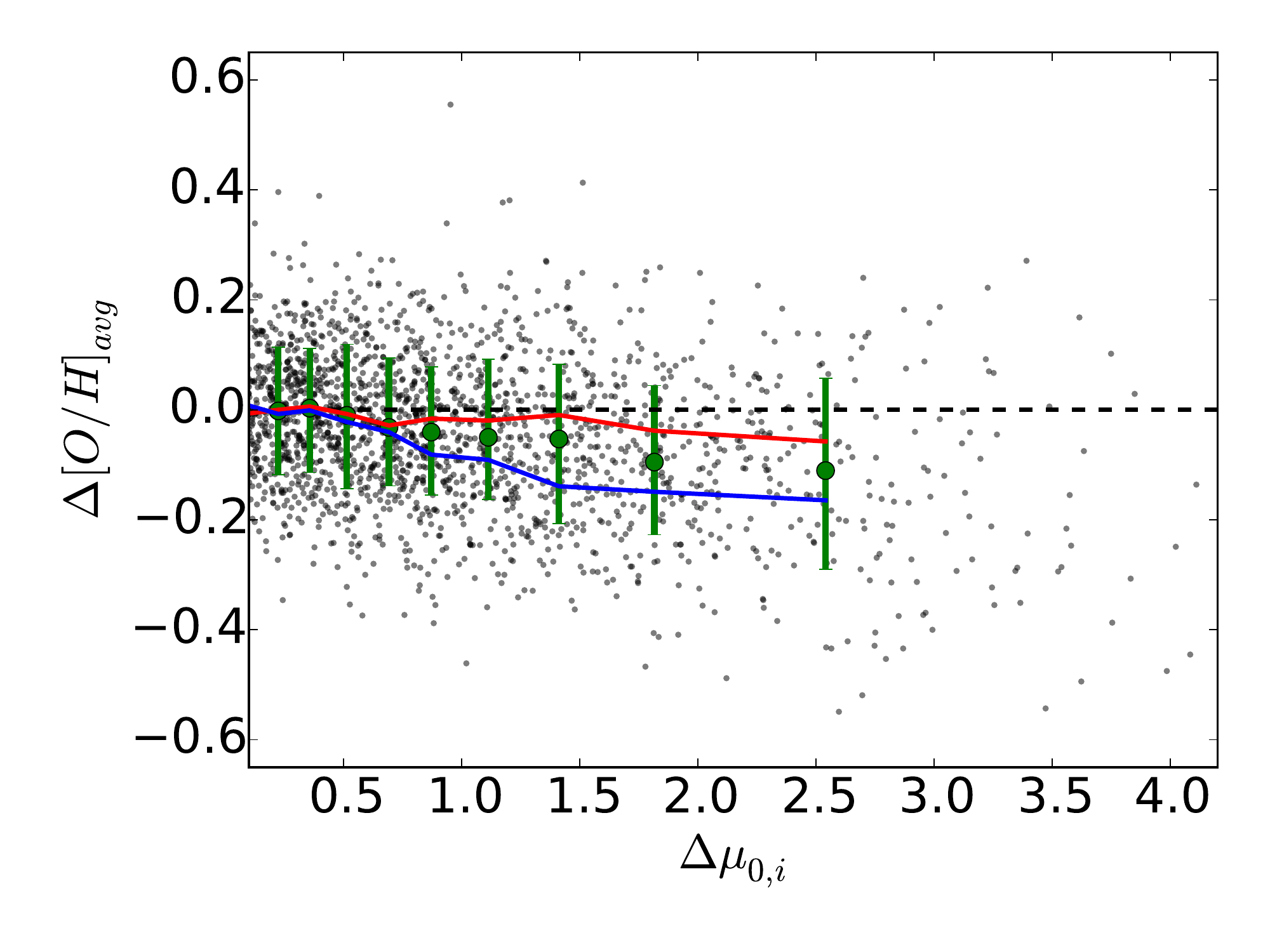}
	\caption{Comparisons between galaxies with similar stellar and gas mass. Each data point comes from two galaxies with $\Delta \log(M_\ast) < 0.3$ and $\Delta \log(M_g) < 0.3$. The horizontal axis is the difference in surface brightness, calculated from the lower surface brightness one (larger $\mu_{0,i}$) minus the higher one (smaller $\mu_{0,i}$). The vertical axis is the difference in average metallicity. We divide the whole data set into 10 equal-size $\Delta \mu_{0,i}$ bins. The median and 16th and 84th percentile of the distribution in each bin is shown as the green error bar. There is a clear trend that, after taking both stellar and gas mass into consideration, lower surface brightness galaxies are on average more metal-poor. We further split the sample into 2 stellar mass bins, and plot the median of each subsample for each $\Delta \mu_{0,i}$ bins in red ($M_\ast > 10^{10.5} M_\odot$) and blue ($M_\ast <= 10^{10.5} M_\odot$), respectively. The surface brightness effect is larger at low mass than in high mass.}
	\label{fig:mz_spec2}
\end{figure}

Next, we show the average metallicity in Figure~\ref{fig:mz_spec}b. Here, at $M_\ast \lesssim 10^{10.5} M_{\odot}$, the distinction between two samples is not as clear as in Figure~\ref{fig:mz_spec}a, but galaxies in the higher surface brightness subsample still tend to locate at the top half of the mass-metallicity relation. We have computed the average metallicity with a few different assumptions of gas profiles (see Section~\ref{sec:model} for profiles used) and the results are similar. By comparing Figure~\ref{fig:mz_spec}a and Figure~\ref{fig:mz_spec}b, we found that, as argued by \citet{ell08}, the combination of abundance gradient and aperture has its effect, but does not account for all the surface brightness dependence. Using the same data set, \citet{pil14b} examined the relation between the disk surface brightness and the metallicity at the center and $R_{25}$. They found that either at the center or $R_{25}$, galaxies with higher surface brightness on average have higher metallicities. This result also supports that the surface brightness of disk has an effect on the metallicity. 

The gas-phase metallicity is the ratio between metals and hydrogen. The nominator, the metal content, is related to the stellar mass, where the denominator is the total gas content of the galaxy. In observation, metallicity has been found to be tightly correlated with both the gas mass and stellar-to-gas mass ratio \citep{bot13,hug13,zah14}. If lower surface brightness galaxies are on average more gas-rich, it would provide a direct explanation of the observed surface brightness dependence. A proper comparison will be between galaxies with both same stellar and gas masses but difference surface brightnesses.

To make such a comparison, we pick out pairs of galaxies with similar stellar and gas mass ($dM < 0.3$ dex) from the sample and compute the difference in surface brightness ($\Delta \mu_{0,i}$) and metallicity ($\Delta [O/H]_{avg}$) between the two galaxies. Figure~\ref{fig:mz_spec2} shows the metallicity deficiency of a lower surface brightness galaxy compared to a higher surface brightness galaxy with similar masses. Each data point represents a comparison between two galaxies. A galaxy can appear more than once in the figure because there can be multiple galaxies with similar stellar and gas mass. We divide the data points into 10 equal-size $\Delta \mu_{0,i}$ bins and show the median and the 16th and 84th percentile of the distribution in each bin. 

The difference, $\Delta [O/H]$, starts from 0 at $\Delta \mu_{0,i} = 0$, and gradually becomes more negative as $\Delta \mu_{0,i}$ increases (lower surface brightness). This result indicates that, at the same stellar \textit{and} gas mass, the metallicity depends on surface brightness of galaxies. The difference between low and high surface brightness galaxies seen in Figure~\ref{fig:mz_spec} cannot be entirely contributed to their gas content. 
We further split the sample into two stellar mass bins. The surface brightness dependence is more prominent in the lower mass bin. The surface brightness effect is presented only at $\Delta \mu_{0,i} \gtrsim 2$ in the higher mass bin, therefore it is not identified in Figure~\ref{fig:mz_spec}.

\subsection{Stellar Metallicity from Broadband Colors}

\begin{figure}[t]
	\includegraphics[width=\columnwidth]{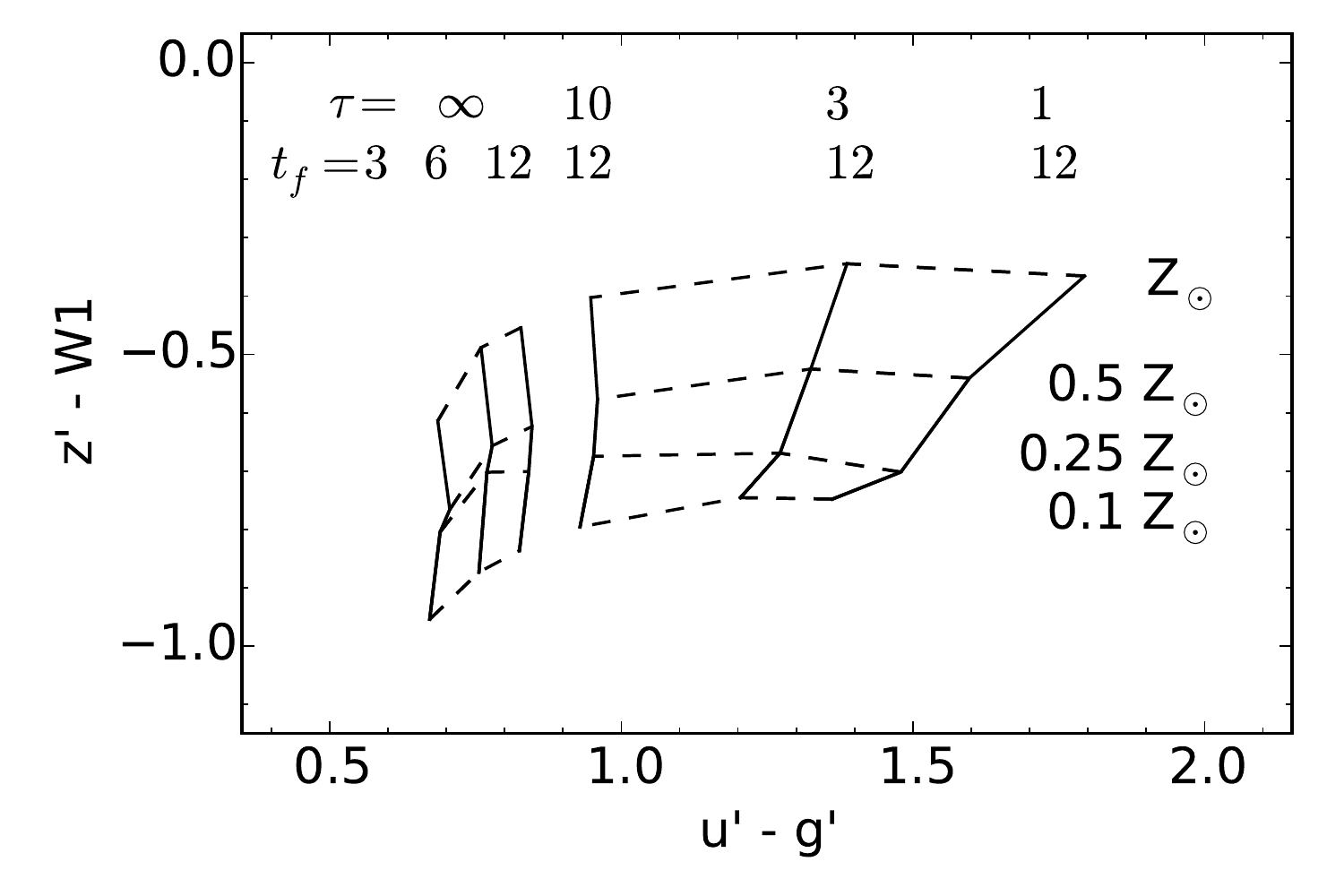}
	\caption{Grids of stellar population synthesis models FSPS for different star formation histories and metallicities. The $\tau$ is the decay time scale of exponential SFH. $\tau = \inf$ indicates constant star formation rate. The $t_f$ (in Gyr) is the time when star formation started. The model shows that the $z'-W1$ color is sensitive to metallicity but insensitive to age for older stellar population. At $u'-g' \gtrsim 0.9$, the $z'-W1$ alone gives a good estimate of stellar metallicity. At $u'-g' \lesssim 0.9$, the $z'-W1$ color still can be used if the $u'-g'$ color is known.}
	\label{fig:model}
\end{figure}

We have seen that the average gas-phase metallicity of a galaxy depends on its surface brightness, and the effect is more prominent at lower masses. It is intriguing to see whether the same dependence also presents for stellar metallicity. Without metallicity measurements from spectroscopy, we use broadband color as a proxy for stellar metallicity. The broadband color of an unresolved population is affected by both stellar age and metallicity. Therefore inferring metallicity from broadband colors alone is non-trivial. Previous studies showed that, instead of optical colors alone, combining optical and NIR colors could break the age-metallicity degeneracy \citep{gal02,mac04,emi08}. Figure~\ref{fig:model} shows stellar population synthesis models of various metallicities and star formation histories generated by the stellar population synthesis code FSPS \citep{con10} on the $z'-W1$ v.s. $u'-g'$ color-color diagram. The $z'-W1$ color is sensitive to metallicity and insensitive to age, except for galaxies dominated by young stellar populations.  At $u'-g' \gtrsim 0.9$, at fixed metallicity, the change in $z'-W1$ color with age is less then 0.1 magnitudes. Even extending to the youngest population, the $z'-W1$ changes by at most $\sim 0.2$ mag. Therefore, the $z'-W1$ color serves as a good proxy for stellar metallicity. Although different models can yield different metallicities from broadband colors as pointed out in previous studies \citep{lee07,emi08}, here we are only interested in relative metallicity changes but not the actual value. The model dependence is then less important.

\begin{figure}[ht]
\includegraphics[width=1\columnwidth]{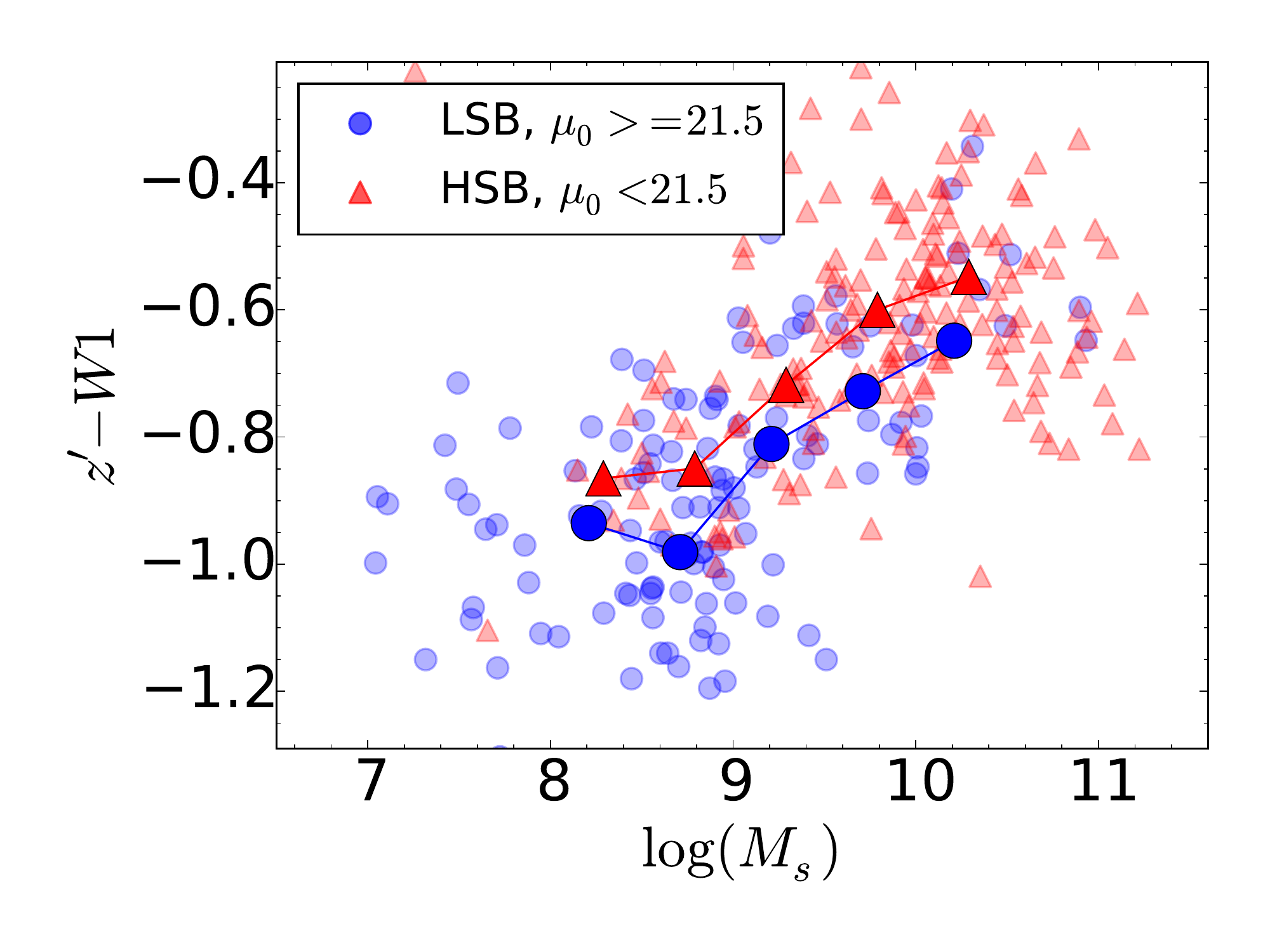}
	\caption{The $z'-W1$ color as a function of stellar mass of the photometric sample. Only galaxies with $u'-g'>0.9$ are shown. The higher and lower surface brightness galaxies are red triangles and blue circles, respectively. We plot the median $z'-W1$ colors in every 0.5 dex $M_\ast$ bin between $8 < \log (M_\ast/M_\odot) < 10.5$, where each subsample contains at least 15 galaxies. The $z'-W1$ color increases as stellar mass increases as expected. Lower surface brightness galaxies on average have more negative $z'-W1$ colors than higher surface brightness galaxies in all stellar mass bins shown. }
	\label{fig:zw1_ms}
\end{figure}

Figure~\ref{fig:zw1_ms} shows $z'-W1$ colors of the photometric sample as a function of stellar mass. Similar to Figure~\ref{fig:mz_spec}, we split the sample into lower and higher surface brightness subsamples. Here we include only galaxies with $u'-g' > 0.9$ to avoid the effect from young populations. We plot the median $z'-W1$ colors of each subsample every 0.5~dex, for $10^{8.0} M_{\odot} \geqslant M_\ast \geqslant 10^{10.5} M_{\odot}$.

\begin{figure}[ht]
	\includegraphics[width=1\columnwidth]{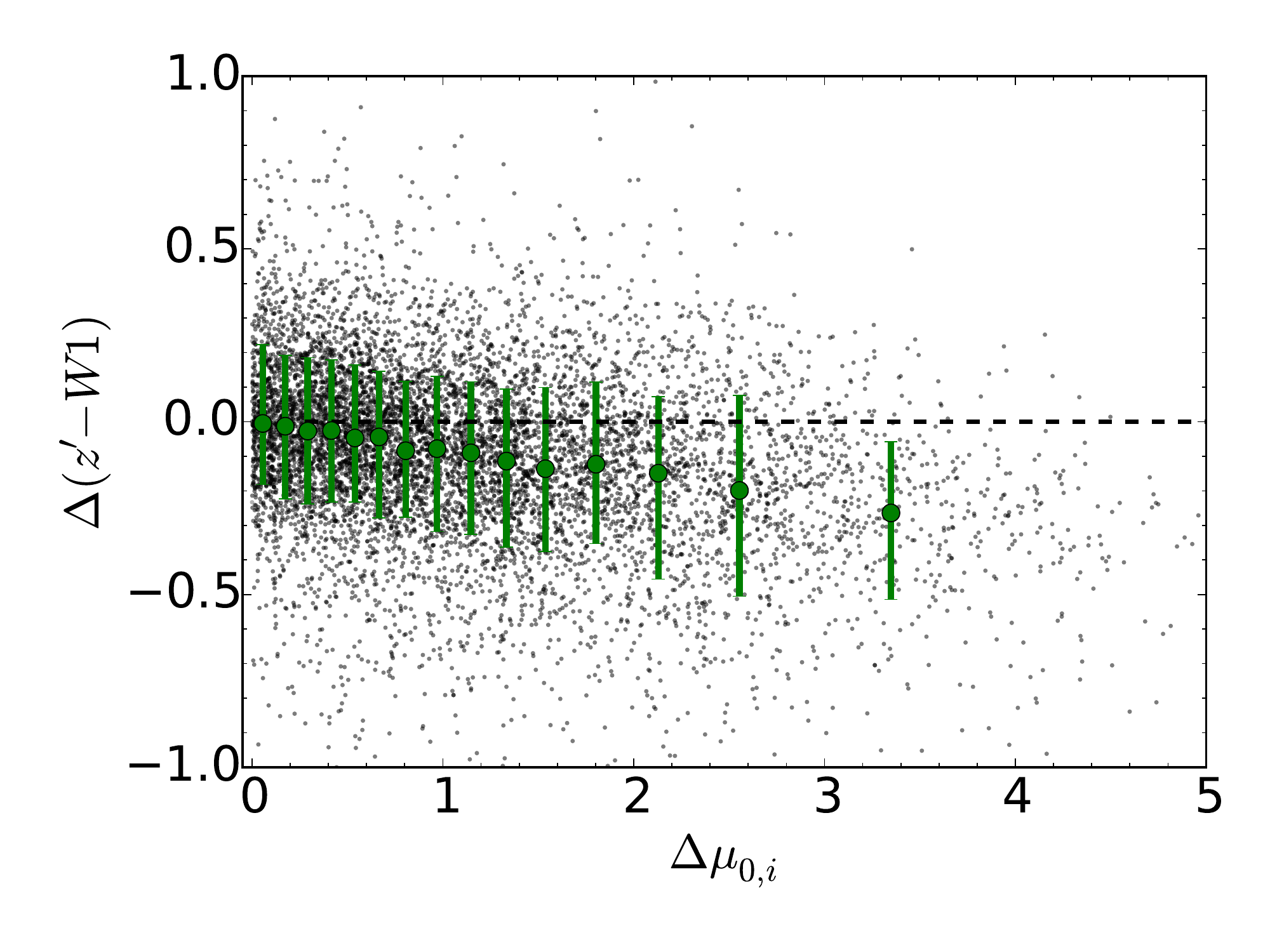}
	\caption{The comparison between galaxies with similar stellar and gas masses. Each data point comes from a pair of galaxies with $\Delta \log(M_\ast) < 0.3$ and $\Delta \log(M_g) < 0.3$. For galaxies with $u'-g'<0.9$, we further require $\Delta (u'-g') < 0.05$. The data set is divided into 15 equal-size $\mu_{0,i}$ bins and the median and 16th and 84th percentile in each bin is shown by the green error bar. At the same stellar and gas masses, lower surface brightness galaxies have smaller $z'-W1$ colors, indicating lower stellar metallicities.}
	\label{fig:dmu_zw1}
\end{figure}

First, overall, more massive galaxies generally have larger $z'-W1$ color, indicating higher stellar metallicity. This conclusion is in line with the result from SDSS fiber spectra \citep{gal05}. Moreover, similar to the gas-phase metallicity, higher surface brightness galaxies also have larger $z'-W1$ colors at given stellar mass. In all 5 stellar mass bins, the high surface brightness subsample is $\sim 0.15$ mags redder than the low surface brightness subsample. Here we examine the effect of surface brightness averaging the whole galaxy. \citet{gon14} discussed the effect of local surface densities on local stellar metallicities using IFU data. They concluded that the local surface brightness regulates metallicity, where denser regions have on average higher metallicities. Our result is consistent with this study. 

Similar to Figure~\ref{fig:mz_spec2}, Figure~\ref{fig:dmu_zw1} compares the $z'-W1$ colors of two galaxies with similar stellar and gas masses. For galaxies with $u'-g' \leq 0.9$, we further require $\Delta(u'-g') \leq 0.05$ in addition to the mass limit in order to mitigate the age effect. We divide the data into 15 equal-size $\Delta \mu_{0,i}$ bins and plot the median, 16th and 84th percentile of the distribution. Again, the difference in color starts from 0 at $\Delta \mu_{0,i} = 0$ and becomes more negative at larger $\Delta \mu_{0,i}$, indicating that lower surface brightness galaxies have systematically lower stellar metallicities for a given stellar \textit{and} gas mass. 

We have estimated the internal extinction following the prescription of \citet{ver01} based on the \ion{H}{1} line width and the inclination of galaxies, which we have the information for all sample galaxies. However, this prescription does not consider the surface brightness of galaxies. If the metallicity of galaxies depends on the surface brightness, the internal extinction would depend on the surface brightness as well. Thus, the prescription of \citet{ver01} could systematically overestimate the internal extinction of low surface brightness galaxies if they are metal-poorer as suggested in previous studies. An arguably better method is to estimate from the flux ratio between total infrared (TIR) and UV luminosities, i.e., the attenuated and un-attenuated light \citep{kon04,bua05,gil07}. However, only $\sim 20\%$ of our galaxies are detected in the W4 band, so this method cannot be applied in our case. When IR data are not available, a common alternative is using UV spectral slope or UV color to infer the TIR-UV flux ratio \citep{cor06,sal07} because there is a positive correlation between the TIR-UV flux ratio and the slope of the UV spectrum \citep[the $IRX-\beta$ relation;][]{kon04,gil07}. But meanwhiles, the correlation is found to depend on ages and dust properties of galaxies \citep{kon04,cor06,gil07}.

To make sure our results are not inherited from the potential systematics of the \citet{ver01} prescription, we also carry out the analysis with internal extinction estimated from the observed UV colors for galaxies with both GALEX FUV and NUV imaging ($\sim 90\%$ of our samples). We follow the calibration of \citet{hao11}:
\begin{equation}
A_{FUV} = 3.83 \times [(FUV-NUV)_{obs} - 0.022],
\end{equation}
and find that both prescriptions for internal extinction give qualitatively similar results. 

We would like to point out that, in both our samples, the surface brightness dependence is only moderate comparing to variations from individual galaxies. The individual variation may dominate in galaxy samples which are too small in size or do not cover a wide enough range of surface brightnesses. Therefore the surface brightness effect may not be detected. It will be possible in the near future to draw a cleaner picture with upcoming integral field unit (IFU) surveys such as MANGA \citep{bun15} or SAMI \citep{bry15}, which can provide spatially resolved metallicity maps of large number of galaxies over wide ranges of mass and surface brightness. 

Also, we note that under the assumption of an exponential disk our result on the surface brightness dependence could be interpreted as a scale length (h) dependence, where galaxies at the same mass with higher central surface brightness have smaller scale lengths.

Also, we note that under the exponential disk assumption our result on the surface brightness dependence could be interpreted as a scale length ($h$) dependence, because galaxies at the same stellar mass with higher central disk surface brightnesses have smaller scale lengths. In general, galaxies are not perfect exponential disks, and this conversion between $\mu_0$ and $h$ will depend on the structure of galaxies. For our cases, we have repeated the similar analysis shown in Figure~\ref{fig:mz_spec2} and Figure~\ref{fig:dmu_zw1}, but as a function of $\Delta h$, and find similar results. 

\section{Theoretical Considerations} 
\label{sec:model}

A large number of studies have modeled the observed mass-metallicity relation with different approaches and formulations. It is widely agreed that the close-box assumption yields too high metallicity. Outflows removing metal-enriched ISM and/or inflows of pristine gas diluting ISM are required to quantitatively match the observed mass-metallicity relation \citep{spi10,pee11,zah12,lil13,zah14}. 

While in their work, the spatially integrated mean relation is successfully reproduced, there is little work addressing the scatter around the mean. We have shown in Section~\ref{sec:result} that, for a given mass, the metallicity is correlated with the surface brightness. It is intriguing to investigate how the spatially resolved surface brightness and metallicity profiles are affected by the inflow and outflow properties.

For this purpose, we apply the chemical evolution model of \citet{kud15}, which addresses the relation among inflow, outflow, metallicity and stellar-to-mass ratio in the spatially resolved case. The inflow and outflow of a galaxy are parameterized by two factors:
\begin{equation}
\Lambda = \frac{\dot{M}_{accr}}{\psi}
\end{equation}
and
\begin{equation}
\eta = \frac{\dot{M}_{loss}}{\psi}
\end{equation}
where $\dot{M}_{loss}$ and $\dot{M}_{accr}$ are the mass-loss and mass-accretion rate, and $\psi$ is the SFR. 

This model makes a few assumptions. First, $\eta$ and $\Lambda$ are assumed to be constant in time. This assumption constrains only the ratio between the SFR and inflow and outflow rate, therefore, all three quantities can be time variables. Second, the inflow gas is free of metals, and the outflow gas has the same metallicity as the ISM at the time launched. Third, the nucleosynthetic yield ($y$) and the stellar mass return fraction ($R$) are both constant. Under these assumptions, the metallicity can be analytically expressed as
\begin{equation}
\label{eq:Zmod}
Z(t) = \frac{y}{\Lambda} \left\{ 1 - \left[ 1 + (1+\frac{\eta-\Lambda}{1-R}) \frac{ M_{\ast}(t) }{ M_g(t) }  \right] ^{-\frac{\Lambda}{(1-R)(1+\frac{\eta-\Lambda}{1-R})}}  \right\}.
\end{equation}
Therefore, the metallicity is determined by $\eta$, $\Lambda$, and the stellar-to-gas mass fraction at a given time. 
If a galaxy has spatially resolved information on stellar mass, gas mass, and metallicity, the three constraints can be used to find the best-fit $\eta$ and $\Lambda$ of the galaxy. We refer readers to \citet{kud15} for detailed derivation and discussion of the model.

We apply the model only to the spectroscopic sample but not the photometric sample. The stellar metallicity inferred from broadband colors can depend on stellar population synthesis models and is less certain for quantitative work \citep{lee07,emi08}. 

For each galaxy in the spectroscopic sample, we have both the metallicity and stellar mass profiles from observations. For the gas distribution, we follow the form in Equation~\ref{eq:model}, and assume an pure exponential gas disk with slope scaled with $R_{25}$. This assumption is supported by observations between 0.2 and 1.0 $R_{25}$, but is not valid for radii outside this range \citep{big12}. Therefore, our nominal range for analysis is limited between 0.2 $R_{25}$ and 1.0 $R_{25}$. For the amount of gas within this range, we do the analysis under two separate assumptions. The first one is distributing the observed amount of gas exponentially from galaxy center to infinity. Therefore, the gas surface density is determined by both $R_{25}$ and the integrated observed gas content. But the validity of this assumption is potentially affected by the gas profile beyond $R_{25}$. Thus, we assume as an alternative a gas surface density profile, which is only constrained by $R_{25}$ \citep{big12}:
\begin{equation}
\frac{\Sigma_{gas}}{\Sigma_{tran}} = 2.1 \times \exp(-1.65\times r / R_{25}), \Sigma_{tran} \simeq 14\; \mbox{M}_{\odot} / pc^2.
\label{eq:model_bie}
\end{equation}
Although the second assumption, determining gas density profile purely from the size of stellar disk, may seem to be less convincing, it is in fact supported by observations \citep{big12}. 
Comparing the two assumptions with the 19 galaxies studied in \citet{kud15}, which have a resolved gas distribution, we find that the first assumption on average yields $\sim 0.17$~dex lower gas surface density, while the second assumption under-predicts gas surface density by $\sim 0.11$~dex within the nominal radius range. As we will find out later, this range of uncertainty is not a concern for our discussion. 

\begin{figure}[ht]
	\includegraphics[width=1\columnwidth]{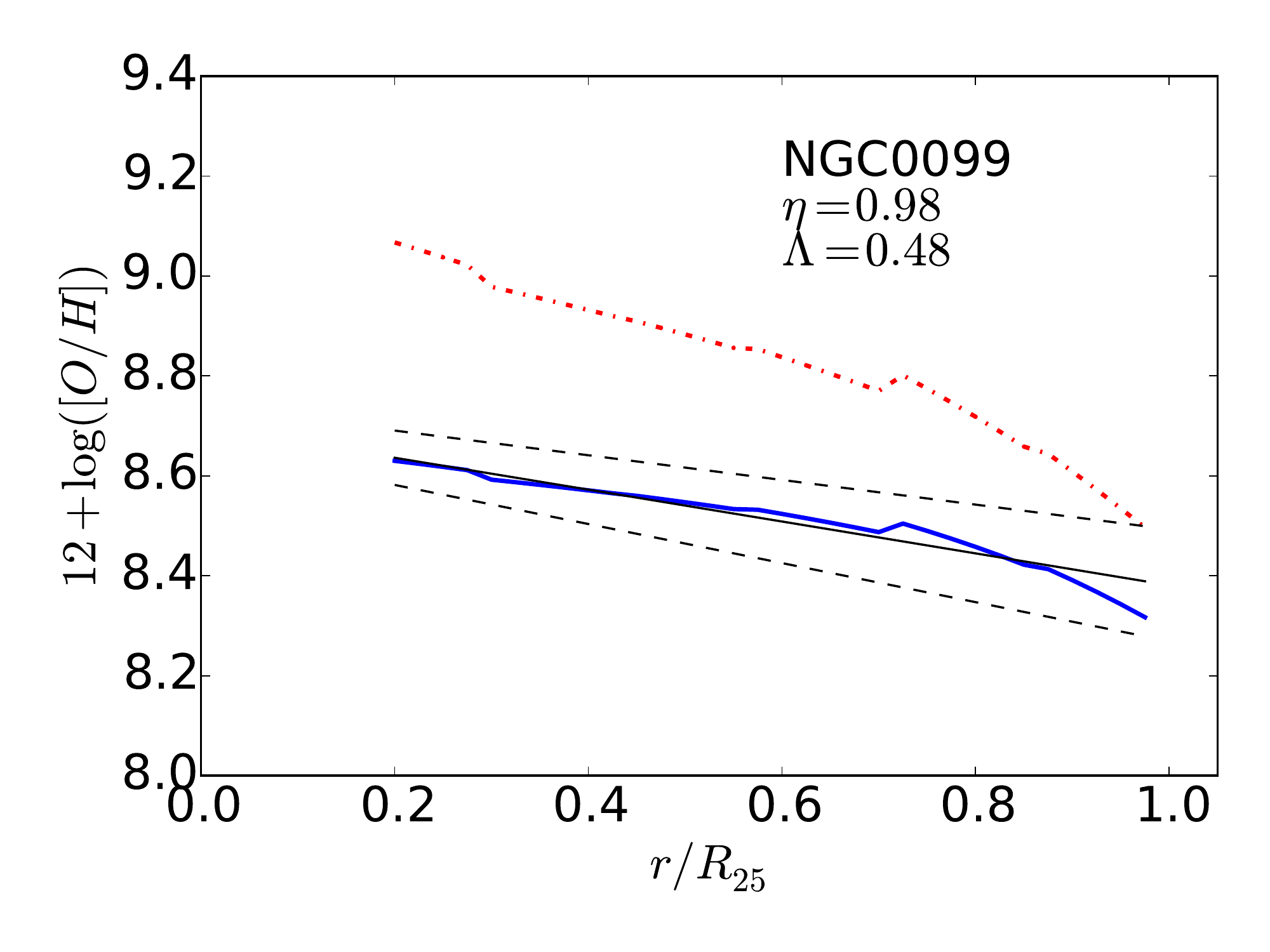}
	\caption{An example of the model fit. The black line represents the observed radial metallicity profile, with the $1 \sigma$ uncertainty indicated by the dashed lines. The best-fit model metallicity profile is plotted as the blue line. The close-box model ($\eta=0, \Lambda=0$) is shown by the red dash-dotted line as a reference.}
	\label{fig:fit}
\end{figure}

For each galaxy we have now the photometrically derived stellar mass column density profiles and gas column density profiles either described by Equation~\ref{eq:model} and the integrated gas mass or by Equation~\ref{eq:model_bie}. We now calculate model metallicity profiles from the stellar-to-gas mass ratio profile and pairs of $\eta$ and $\Lambda$ between 0 and 3 for every hundredth interval. We adopt $R=0.4$ and $y=0.00313$ as calibrated in \citet{kud15}. A 0.15~dex is added to the \ion{H}{2}-region metallicity measured by \citet{pil14a} to match the metallicity zero point obtained by \citet{kud15} as the result of stellar spectroscopy in spiral galaxies. We then compare the model metallicity profile with the observed metallicity profile and compute the $\chi^2$ from the difference between two profiles and the uncertainties for every 0.05 $R_{25}$ from 0.2 $R_{25}$ to 1.0 $R_{25}$ or the last reliable measurement of surface brightness for adopted pairs of $\eta$ and $\Lambda$. The pair of $\eta$ and $\Lambda$ with the minimum $\chi^2$ is then defined as the best fit model. Figure~\ref{fig:fit} shows an example of our comparison between the model and the observation. We note that the assumption on gas profile is valid only in a statistical sense. There is $\sim 60\%$ scatter among individual galaxies. Therefore, we will only discuss the average properties from the model.

\begin{figure*}[ht]
	\includegraphics[width=1\textwidth]{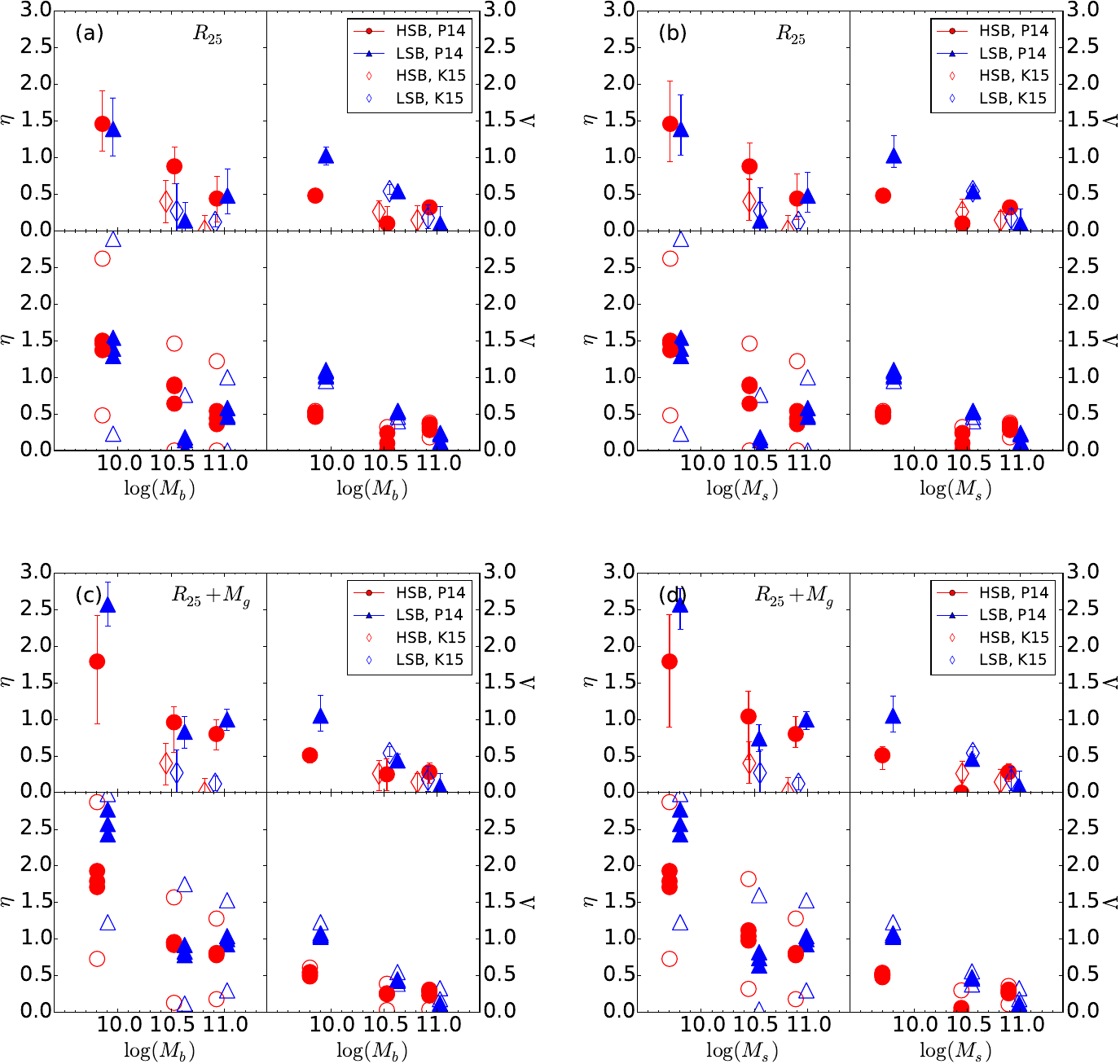}
	\caption{Best-fit $\eta$ and $\Lambda$ as a function of mass and surface brightness. \textit{(a)}: gas amount scaled by $R_{25}$, binned by baryonic mass. \textit{(b)}: gas amount scaled by $R_{25}$, binned by stellar mass. \textit{(c)}: gas amount scaled by $R_{25}$ and measured gas mass, binned by baryonic mass. \textit{(d)}: gas amount scaled by $R_{25}$ and measured gas mass, binned by stellar mass. Galaxies in the spectroscopic sample are first sorted into 3 mass bins then two surface brightness bins in each mass bin. Higher surface brightness galaxies are in red and lower surface brightness galaxies are in blue. In the upper panel of each subplot, the error bars represent for the 16th and 84th percentile of the distribution of $\eta$ and $\Lambda$ in each subsample. The filled symbols in lower panel of each subplot show the uncertainties from varying the slope of surface density of gas. Open symbols illustrate the uncertainties from varying the gas amount (see Section~\ref{sec:model}). The 19 galaxies studied by \citet{kud15} are also plotted as a comparison. }
	\label{fig:eta}
\end{figure*}

Our result with two different gas profiles is presented in Figure~\ref{fig:eta}. Galaxies are first sorted into 3 bins in baryonic (Figure~\ref{fig:eta}a and c) or stellar mass (Figure~\ref{fig:eta}b and d). In each mass bin, galaxies are further separated into two equal-size bins by their surface brightness. Each bin contains $\sim 20$ galaxies. We show the median and the 16 and 84 percentile of the distributions of $\eta$ and $\Lambda$ in each sub-sample. The difference between the median surface brightnesses of two sub-sample is $\sim 1$ mag for the high and mid-mass bins, and $\sim 2$ mags for the low-mass bin. As a comparison, we also plot the best-fit $\eta$ and $\Lambda$ from the 19 galaxies with measured stellar-to-gas mass profiles studied by \citet{kud15}. 

To investigate the uncertainty resulting from the parameterized gas profile, we considered four more different gas profiles. Following the $1 \sigma$ scatter of the slope of the universal gas profile given in \citet{big12}, we first consider two cases where the gradient of the gas distribution is -1.78 or -1.49, instead of the canonical value of -1.65. For the other two cases, we fix the gradient, but double or half the amount of gas in order to access the uncertainty from the assumed amount of gas and the effect from possible systematics (see Section~\ref{sec:sys} for more discussion). For all four cases, the median $\eta$ and $\Lambda$ of each sub-sample is calculated and plotted in lower panels of Figure~\ref{fig:eta}. Variation in slope is shown by light filled symbols and variation in gas amount is shown by open symbols.

Generally, both $\eta$ and $\Lambda$ depend on the mass of galaxy, where higher mass galaxies have on average lower $\eta$ and $\Lambda$. However, $\eta$ depends strongly on the gas amount assumed. Without knowing the amount of gas, $\eta$ is essentially not constrained in the lowest mass and lower surface brightness bin. In all subsamples, the difference in $\eta$ between high and low surface brightness bins could come from systematically different assumed gas amount, if there is any. On the other hand, $\Lambda$ is more robust against variations in gas profiles. In the two low mass bins, lower surface brightness galaxies have larger $\Lambda$ regardless of gas profile assumed. 

The same result also can be seen in galaxies with observed stellar-to-gas mass profiles (open diamonds). Only $\Lambda$ in the low mass bin shows a surface brightness dependence, while $\eta$, and $\Lambda$ in the high mass bin does not seem to depend on surface brightness. However we note that this result comes from a small sample of only 5 galaxies in each bin.

\section{Discussion}

\subsection{Implications from the Mass and Surface Brightness Dependence on $\Lambda$}

Chemical evolution model suggests that, for given mass, lower surface brightness galaxies tend to have larger $\Lambda$. In this section we discuss possible explanations for the correlation between $\Lambda$ and the surface brightness of a galaxy.

\subsubsection{Star Formation Efficiency}

The accretion parameter $\Lambda$ is the gas accretion rate divided by star-formation rate. It can be understood as the reciprocal of star formation efficiency; while gas is falling onto the galaxy, $\Lambda$ monitors how much of it is turned into stars. A larger $\Lambda$ indicates inefficient star formation. 

Our model suggests that lower surface brightness galaxies are less efficient in turning gas into stars, which is in agreement with the prevalent idea. Observations of giant low surface brightness galaxies ($M_\ast \gtrsim 10^{10} M_{\odot}$) showed that they have copious amounts of \ion{H}{1} gas that is comparable to their stellar components \citep{mat01,one04}. Their gas-to-stellar mass ratio is higher than their high surface brightness counterparts \citep{bur01,one07}. By comparing star formation rate and \ion{H}{1} mass, \citet{boi07} found that low surface brightness galaxies have lower SFR compared to high surface brightness galaxies with similar \ion{H}{1} mass, evidence that low surface brightness galaxies are inefficient in star formation. 

The low star formation efficiency in low surface brightness galaxies is also suggested by theory that low surface brightness galaxies are stable against both global and local growth of instability \citep{dal97,mih97,may04,gho14}. Stability analysis on individual low surface brightness galaxies demonstrated that they have high Toomre $Q$ parameter \citep{mih97,gho14}, where $Q \equiv \sigma\; \kappa/3.36\; G\; \Sigma > 1$ for stability, and $\sigma$ and $\kappa$ are the radial velocity dispersion of the stars and the epicyclic frequency, respectively \citep{too64}. 

Structures like spiral arms and bars are also harder to form due to a lack of global instability \citep{mih97}. Gas clouds in low surface brightness disks experience less cloud collisions and compression, which can trigger star formation. This proposition is also supported by the fact that fewer bars are seen in low surface brightness galaxies \citep{imp96}. The larger $\Lambda$ inferred from our model, thus, agrees with the majority of previous studies in terms of star formation efficiency.

\subsubsection{Environmental Effect}

Since $\Lambda$ is the ratio between mass accretion rate and star formation rate, for two galaxies with the same stellar mass, the one with higher $\Lambda$, i.e. lower surface brightness, should have had more inflow material over time. This may imply that low surface brightness galaxies reside in higher density region than higher surface brightness galaxies of the same mass, or more likely to be the central galaxy in a halo because central galaxies have overall higher accretion rates shown by hydrodynamical simulations \citep{ker09}. 

This assertion seems to be at odds with the general perception, that low surface brightness galaxies have fewer neighboring galaxies on scales of $\lesssim$ 1~Mpc to several Mpc \citep{bot93,ros09,gal11}. But this is not necessary in conflict with our results. 
While there are several ways to define the surface brightness of a galaxy, low surface brightness galaxies in the literature refer mostly to galaxies whose surface brightness are fainter than a certain value. This definition yields a heterogeneous sample. A galaxy with low surface brightness could be a result of an extended distribution of baryons, its less massive nature, or being massive but gas-rich (star-poor) condition. Each type of galaxies may have its own environmental dependence, and the contribution from each type is not clear. 

\citet{ros09} found that low surface brightness galaxies reside in underdensed regions at scales $\gtrsim 2$~Mpc from SDSS DR 4 sample. In the same study, they further split the sample into two redshift bins and showed that this environmental dependence is more prominent in the low-$z$ bin than in the high-$z$ bin. As shown in \citet{ros09}, their low-$z$ low surface brightness sample consists mainly of dwarf galaxies, and the high-$z$ low surface brightness sample is dominated by regular spirals. The behavior in two redshift bins may suggest that, at least part of the apparent environmental effect is inherited from the luminosity, or mass, of the galaxy. In Figure~\ref{fig:eta} we are comparing galaxies with the same mass, where the difference should reflect the effect from galaxy structure but less from mass. Also, \citet{bla05} studied the relationship between environment and optical properties of galaxies in the SDSS and found that, lower surface brightness galaxies are in denser region for a given luminosity. The result from \citet{bla05} and our result suggest that it would be necessary to investigate each type of galaxies separately and not put heterogeneous individuals under the same umbrella. 

Although our result indirectly suggests that galaxies with more extended disks tend to reside in denser regions, we realize at the same time that the question whether the environment affects galaxy structure is beyond the scope of this paper and worth its own dedicated study. 

\subsubsection{Assembly History}

The discussion so far is under the assumption of constant $\Lambda$. This assumption has the physical motivation that inflow increases a gas reservoir therefore the capability of forming stars, but it is hard to verify observationally. Some observations of high-redshift massive galaxies ($1<z<3$) suggested $\Lambda > 1$ \citep{tac10,tac13,yab15}, which is larger than what we derive for local galaxies ($\Lambda < 1$) therefore constitutes a challenge to our assumption of constant $\Lambda$. Nevertheless, it is not clear whether galaxies in different studies can be directly compared, i.e., our sample is not necessary the descendant of galaxies in previous studies. Moreover, in spite of possible flaws in the assumption, this model has been shown to be able to reproduce the distribution of metallicity gradients of local galaxies \citep{ho15}. Practically, the metallicity is not sensitive to inflow and outflow in the gas-rich regime. Because metallicity is the content of metals normalized by the amount of gas, inflow and outflow would not drastically change it when the galaxy still possess a vast gas reservoir. Therefore, the model parameters, $\eta$ and $\Lambda$, reflect the more recent accretion activities and less the early phase of galaxy formation. Although the assumption of constant $\eta$ and $\Lambda$ may not necessarily hold, we consider our fitting parameters should at least reflect the inflow and outflow properties at recent times.

If we assume that $\Lambda$ is decreasing over time, as hinted by observations, the dependences on surface brightness we see from local galaxies are qualitatively consistent with the canonical view of halo formation. More compact galaxies are considered to reside in compact halos with lower spin which form relative early \citep{jim98,mo98}. 
For baryons, hydrodynamic simulation showed that most baryons in low spin halos collapse into the center of the galaxies within the first $\sim 1.5$~Gyrs \citep{kim13}, therefore implying low accretion at the present day. 

Overall, galaxies with higher surface brightness have likely passed their peak accretion phase, while low mass, lower surface brightness galaxies are still assembling their baryons. The low $\Lambda$ observed in the high mass bin could also result from the virial shocks developing in halos with $M_{halo} \gtrsim 10^{12} M_{\odot}$, which suppress cold material falling onto the galaxies \citep{ker05,ker09,bro09}. In this case, the mass is the dominant factor, and the surface brightness effect is relatively small and, therefore, not observed. Both factors could lower the accretion rate, $\dot{M}_{accr}$ in high mass, high surface brightness galaxies. 
On the other hand, it has been shown that low surface brightness galaxies on average have lower SFR \citep{one07,hua12}, which affects the denominator of $\Lambda$. Therefore, low mass, low surface brightness galaxies would have an average higher $\Lambda$, from the combination of stronger accretion and lower SFR.

\subsection{Possible Systematics}
\label{sec:sys}

\subsubsection{Molecular gas content}

In Section~\ref{sec:result}, we compare galaxies with the same stellar and gas mass, where molecular mass is estimated from a stellar-mass-dependent $M(H_2)/M(HI)$ fraction from \citet{bot14}. Under this assumption, galaxies with the same stellar and HI mass will have the same amount of molecular gas. However, one could naturally expect that lower surface brightness galaxies have a deficit in molecular gas due to their lower metallicity, so that the neutral gas will not be able to cool enough to form H$_2$. 

Detections of molecular gas in low surface brightness galaxies suggested that they have low $M(H_2)/M(HI)$ fraction compared to galaxies with higher surface brightness \citep{one03,one04,mat05,das06}. As a result, our estimate could lead to systematic overestimate of total gas content in lower surface brightness galaxies. 

Based on the result of \citet{bot14}, the scatter of the $M(H_2)/M(HI)$ fraction around the mean value is $\sim 0.4$~dex for given stellar mass. Some of the scatter is possibly inherited from the correlation with metallicity and surface brightness. According to Equation~\ref{eq:mol}, the H$_2$ mass is less than 10\% of \ion{H}{1} at $M_\ast \lesssim 10^{10} M_\odot$. The 0.4~dex scatter corresponds to $\lesssim 25\%$, or $\lesssim 0.1$~dex, of total gas mass, and is therefore negligible. 

Of more concern is the high mass end, where $M_\ast \simeq 10^{11}M_{\odot}$. At this stellar mass, the average H$_2$ mass is comparable to \ion{H}{1} mass, and the scatter in total gas mass is $\sim 0.25$~dex. To account for this possible systematic effect, we have allowed a 0.3~dex difference in mass for galaxies to be considered as ''the same mass'' in Section~\ref{sec:result}. 
Moreover, for our chemical evolution model, we have run the test cases of different gas content in Section~\ref{sec:model} and shown the result in Figure~\ref{fig:eta}. Our discussion on $\Lambda$ is not affected by changing the gas content within the nominal range. Especially at low mass, where the surface brightness effect is more prominent, the molecular gas content is essentially negligible. 

\subsubsection{Metallicity scale}
In Section~\ref{sec:model}, we add a correction of 0.15~dex to the metallicities measured by \citet{pil14a}. This correction has been found by \citet{kud15} and is the result of a comparison of metallicities obtained from detailed quantitative spectroscopy of blue supergiant stars with \ion{H}{2}-region metallicities measured by \citet{pil14a}. The yield ($y = 0.00313$) used in our chemical evolution model of section~4 has been empirically determined by \citet{kud15} from the metallicity and metallicity gradient in the Milky Way disk as observed from Cepheids and B-stars The yield and the metallicity scale used in Section~\ref{sec:model} are,thus, consistently based on quantitative stellar spectroscopy.

\begin{figure*}[t]
	\includegraphics[width=0.99\textwidth]{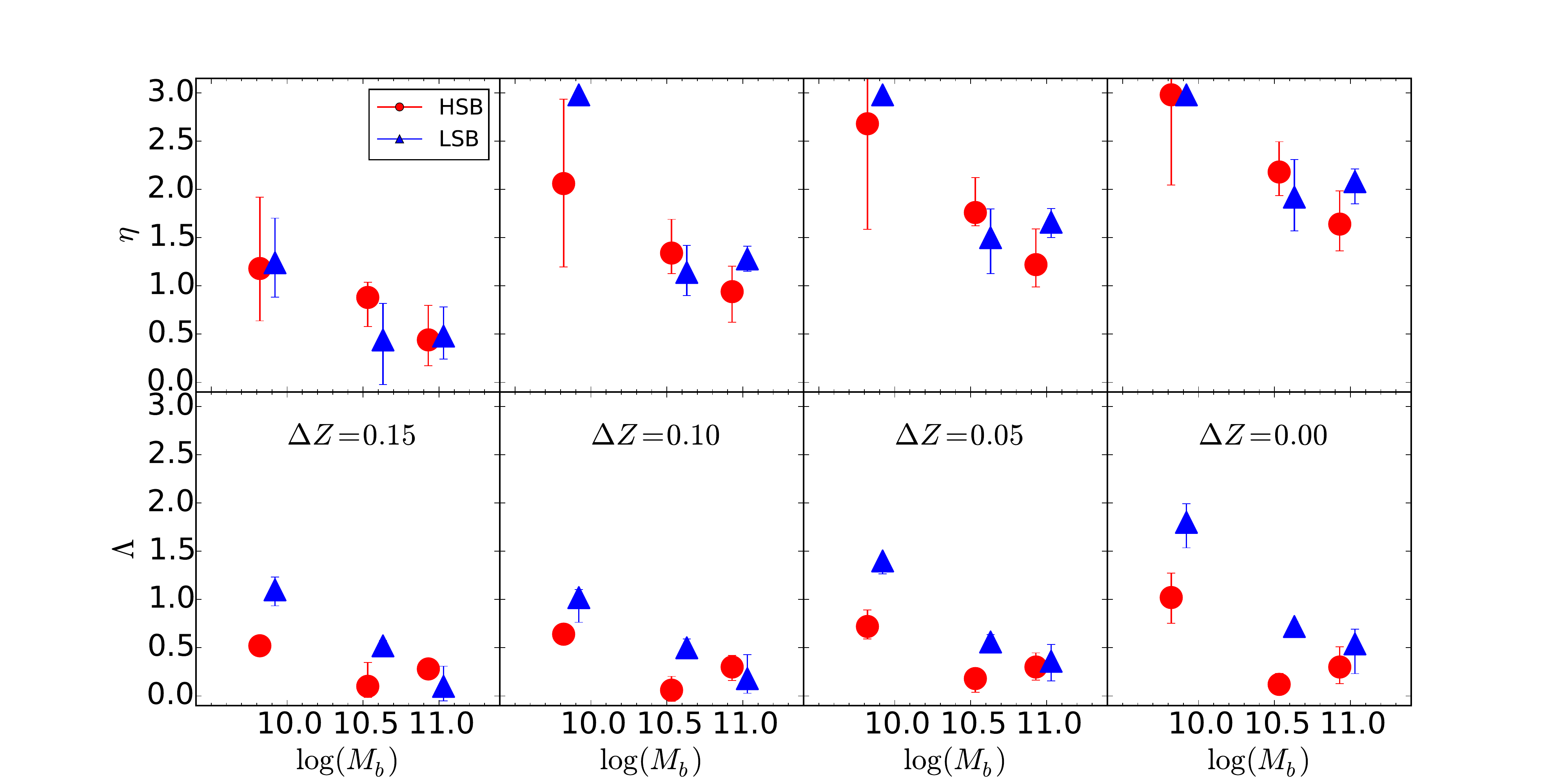}
	\caption{Test of the systematical effect from the zero point of metallicity scale on $\eta$ and $\Lambda$. We adopt 4 different metallicity scales, adding 0.15~dex (the scale adopted for this work), 0.10, 0.05, and 0.00~dex to the metallicities of \citet{pil14a}. Gas profiles are scaled by $R_{25}$ as in Figure~\ref{fig:eta}a. The zero point affects $\eta$ and $\Lambda$ in a quantitative sense, but not our conclusion, where lower surface brightness galaxies have higher $\Lambda$ values than higher surface brightness galaxies of similar masses. }
	\label{fig:eta_test}
\end{figure*}

The inflow and outflow strength determined in our model could be affected by systematic uncertainties of the metallicity scale used. To access this uncertainty we apply a similar procedure as \citet{kud15} We repeat the analysis of Section~\ref{sec:model}, but this time adding corrections to the \citet{pil14a} metallicities of 0.10, 0.05, and 0.00~dex, respectively. Figure~\ref{fig:eta_test} shows the effects from changes of the zero point. While the zero point is important in a quantitative sense (decreasing the metallicity leads to larger values of $\eta$ and $\Lambda$), it does not affect our major conclusion. Lower surface brightness galaxies have higher $\Lambda$ values than higher surface brightness galaxies of similar mass. In Figure~\ref{fig:eta_test} we only show the results for one type of gas profiles and only binned by baryonic mass corresponding to the case in Figure~\ref{fig:eta}a. Application of the zero point shifts to the other cases lead to a similar conclusion. 

We note that changing the zero-point of the metallicity scale is equivalent to changing the yield in the chemical evolution model, as can be seen in Equation~\ref{eq:Zmod}. The result discussed here can also be seen as a test of the influence from the uncertainty of the yield.

\section{Summary}

In this paper, we investigate whether the metallicity of galaxies depends on their surface brightnesses and discuss the possible origin of the dependence. 

\begin{itemize}
	\item[1.] Previous studies found that, at given stellar mass, low surface brightness galaxies are more metal-poor. These studies used SDSS fiber spectra, where the aperture effect is difficult to account for. We show that the result remains valid when the average metallicity of the whole galaxy is considered. 
	\item[2.] The surface brightness dependence still exists if gas mass is also taken into account. For galaxies with similar stellar and gas masses, lower surface brightness systems tend to have lower metallicity. 
	\item[3.] We use chemical evolution models to constrain the inflow and outflow properties of galaxies. The ratio between accretion rate and star formation rate ($\Lambda$) is larger in low mass, lower surface brightness galaxies. The surface brightness effect is more prominent in low mass galaxies. On the other hand, the outflow property is not well constrained. 
	\item[4.] The high $\Lambda$ in lower surface brightness galaxies can be directly interpreted as low star formation efficiency. This interpretation is consistent with the prevalent idea.
	\item[5.] For galaxies with the same stellar mass, lower surface brightness galaxies should have more inflow material overtime because of higher $\Lambda$, which is the ratio between mass accretion rate and star formation rate. This result indirectly suggests that lower surface brightness galaxies reside in denser regions than higher surface brightness galaxies of the same mass, or more likely to be the cetral galaxy in a halo, therefore they have more inflow material. 
	\item[6.] If $\Lambda$ is not constant in time as assumed in the model, the surface brightness dependence on $\Lambda$ may be an indication of different accretion histories, where low mass, lower surface brightness galaxies are currently at their major accretion phase, while this phase in high mass, higher surface brightness galaxies has passed. 
\end{itemize}

In the near future, large IFU surveys will provide maps of metallicity, and perhaps measurements of inflow and outflow activities, in a large number of galaxies spanning wide ranges of galaxy properties. We will understand better how galaxy structure affects the evolution of galaxies with the aid of IFU data. 

\acknowledgments
RPK acknowledges support by the National Science Foundation under grants AST-1108906 and AST-1008798. RBT acknowledges support from the US National Science Foundation award AST09-08846 and NAA award NNX12AE70G.
This publication makes use of data products from the \textit{Wide-field Infrared Survey Explorer (WISE)}, which is a joint project of the University of California, Los Angeles, and the Jet Propulsion Laboratory/California Institute of Technology, funded by the National Aeronautics and Space Administration. We acknowledge the usage of the HyperLeda database (http://leda.univ-lyon1.fr). This work makes use of data products from the SDSS. Funding for SDSS-III has been provided by the Alfred P. Sloan Foundation, the Participating Institutions, the National Science Foundation, and the U.S. Department of Energy Office of Science. The SDSS-III web site is http://www.sdss3.org/. PFW acknowledges the writing retreat of IfA students and postdocs, SWOOP, where part of the manuscript is written.

\end{document}